\documentclass[12pt,epsf]{article}
\usepackage{amsmath,graphicx,epsfig,latexsym}
\usepackage{hyperref}
\usepackage{lmodern}
\usepackage{amssymb}

\numberwithin{equation}{section}

\newcommand{\be}{\begin{equation}}
\newcommand{\ee}{\end{equation}}
\newcommand{\la}{\langle}
\newcommand{\ra}{\rangle}
\newcommand{\beqa}{\begin{eqnarray}}
\newcommand{\eeqa}{\end{eqnarray}}
\newcommand{\nn}{\nonumber}

\newcommand{\cov}{\mathop{{\rm cov}}}
\newcommand{\var}{\mathop{{\rm var}}}

\def\boxit#1{\vbox{\hrule\hbox{\vrule\kern8pt
\vbox{\hbox{\kern8pt}\hbox{\vbox{#1}}\hbox{\kern8pt}}
\kern8pt\vrule}\hrule}}
\def\mathboxit#1{\vbox{\hrule\hbox{\vrule\kern8pt\vbox{\kern8pt
\hbox{$\displaystyle #1$}\kern8pt}\kern8pt\vrule}\hrule}}

\def\IB{\relax\hbox{$\inbar\kern-.3em{\rm B}$}}
\def\IC{\relax\hbox{$\inbar\kern-.3em{\rm C}$}}
\def\ID{\relax\hbox{$\inbar\kern-.3em{\rm D}$}}
\def\IE{\relax\hbox{$\inbar\kern-.3em{\rm E}$}}
\def\IF{\relax\hbox{$\inbar\kern-.3em{\rm F}$}}
\def\IG{\relax\hbox{$\inbar\kern-.3em{\rm G}$}}
\def\IGa{\relax\hbox{${\rm I}\kern-.18em\Gamma$}}
\def\IH{\relax{\rm I\kern-.18em H}}
\def\IK{\relax{\rm I\kern-.18em K}}
\def\IL{\relax{\rm I\kern-.18em L}}
\def\IP{\relax{\rm I\kern-.18em P}}
\def\IR{\relax{\rm I\kern-.18em R}}
\def\IZ{\relax\ifmmode\mathchoice
{\hbox{\cmss Z\kern-.4em Z}}{\hbox{\cmss Z\kern-.4em Z}}
{\lower.9pt\hbox{\cmsss Z\kern-.4em Z}} {\lower1.2pt\hbox{\cmsss
Z\kern-.4em Z}}\else{\cmss Z\kern-.4em Z}\fi}

\def\II{\relax{\rm I\kern-.18em I}}

\begin{document}
\pagestyle{empty}%

\begin{titlepage}
\title{A Priori Tests for the MIXMAX Random Number Generator\footnote{This is an expanded version of the talk given by S. Konitopoulos in Athens meeting of the MIXMAX collaboration  in September 2016 \url{https://indico.cern.ch/event/558996/}. As interesting new results of L'Ecuyer, Wambergue, Bourceret 
have become available to us \cite{L?Ecuyer:2017}, we think that it is worth to share our initial investigation 
on the spectral index which has been performed by a different method. }}
\maketitle
\begin{center}
\author{Spyros Konitopoulos\footnote{spykoni@inp.demokritos.gr} }\\
\author{Konstantin G. Savvidy\footnote{k.savvidis@cern.ch} }
\begin{abstract}
We define two a priori tests of pseudo-random number generators for the class of linear matrix-recursions. The first desirable property of a random number generator is the smallness of serial or lagged correlations between generated numbers. For the particular matrix generator called MIXMAX, we find that the  serial correlation actually vanishes. Next, we define a more sophisticated measure of correlation, which is a multiple correlator between elements of the generated vectors. The lowest order non-vanishing correlator is a four-element correlator and is non-zero for lag $s=1$. At lag $s \ge 2$, this correlator again vanishes. For lag $s=2$, the lowest non-zero correlator is a six-element correlator. The second desirable property for a linear generator is the favorable structure of the lattice which typically appears in dimensions higher than the dimension of the phase space of the generator, as discovered by Marsaglia. We define an appropriate generalization of the notion of the spectral index for LCG which is a measure of goodness of this lattice to the matrix generators such as MIXMAX and find that the spectral index is independent of the size of the matrix N and is equal to $\sqrt{3}$.
\end{abstract}
\end{center}
\end{titlepage}

\newpage
\pagestyle{plain}
\section{Introduction}
We want to study the correlation and spectral properties of the 
recursive matrix random number generator MIXMAX, which is defined by the automorphism of the torus \cite{Savvidy:1991}:
\beqa\label{mixmax uni}
u_{i}(t+1)&=& \sum_{j=1}^{N}A_{ij}u_{j}(t) \bmod1,~~~t=0,1,2,...\nn\\
u_{i}(q)&=&u_{i}(0).
\eeqa
where $u \in [0,1)]$ and $A$ is a specific unimodular matrix with integer elements:
 \beqa
 A&=& \left(   \begin{matrix}
    2 & 3~ & 4 & \dots & N-4 & N-3 & N-2 & N-1 & N & 1 \\
    1 & 2~ & 3 & \dots & N-5 & N-4 & N-3 & N-2 & N-1 & 1 \\
    1 & 1~ & 2 & \dots & N-6 & N-5 & N-4 & N-3 & N-2 & 1 \\
    \dots \\
    \dots \\
    \dots\\
    1 & 1~ & 1 & \dots & 1 & 1 & 2 & 3 & 4 & 1 \\
    1 & 1~ & 1 & \dots & 1 & 1 & 1 & 2 & 3 & 1 \\
    1 & 1~ & 1 & \dots & 1 & 1 & 1 & 1 & 2 & 1 \\
    1 & 1~ & 1 & \dots & 1 & 1 & 1 & 1 & 1 & 1 \\
   \end{matrix} \right).
\eeqa

If the eigenvalues of the matrix $A$ are all different by absolute value from one, then the sequence defines a
deterministic Kolmogorov-Anosov K-system with strong chaotic properties \cite{Savvidy:1991, Savvidy:1991a}.
In what follows we use the fact K-mixing assures that spatial averages equal time averages (ergodicity). Some of the concepts below are borrowed from the theory of stochastic processes of Markov, Wiener etal, keeping in mind that strictly speaking we are applying them to a deterministic process.

\section {The Serial Correlation Test for MIXMAX}

\subsection{The cross-correlation matrix between elements of the pseudo-random vector}

Our first goal is to determine the correlation between separate components of the MIXMAX sequence of generated vectors.

To measure the amount of dependence between $u_{j}(t+s)$ and $u_{i}(t)$, we define the lag-s
covariance of the sequence as follows: 
\be
\cov (u_{i},u_{j}^{s}) = {\la u_{i}(t) ~, ~ u_{j}(t+s) \ra}
\ee
and the correlation matrix as
\beqa
C^{s}_{ij}&=&
{\la u_{i}(t),u_{j}(t+s) \ra} \over {\la u_{i}(t) u_{i}(t) \ra}
\eeqa
where the bracket $\la \ra$ denotes averaging over time $t$.

\subsection {Total correlation}
Another measure of correlation can be derived if, instead of focusing
on the amount of dependence between the individual components of the vectors, we look at the total amount of
dependence between the vectors as a whole. 

The total lag-s covariance is:
\be
\cov  (U,U_{+s} ) = \la U(t) U(t+s) \ra = Tr\Big[\cov\{u_{i},u_{j}^{s}\}\Big]
\ee
so that the total covariance is the trace over $(i,j)$ of the element-wise covariance.

Finally, the total correlation coefficient is:
\be
C_{+s} = { {\la U(t) U(t+s) \ra} \over {\la U(t) U(t) \ra} }
\ee
which turns out to equal to the trace of the correlation matrix, taking into account the fact that the variance of all the components is the same, as we shall see below.

\subsection {Calculation of the  cross-correlation matrix}
From the point of view of a user of a pseudo-random number generator, it is desirable to minimise the correlation between the generated numbers, because an easily detectable correlation immediately contradicts the hypothesis of randomness. In later sections we will define other properties of pseudo-random number generators, which distinguish the pseudo-random numbers which come from a linear generator such as MIXMAX, from physical random numbers.

As a first step
we will calculate the correlation coefficients.
When the period of the sequence is taken to infinity, $q\rightarrow \infty$, all the quantities in the above formulae tend to a finite limit and can be evaluated by replacing the averaging over time with spacial averaging.
\subsubsection {$1D$ case}
Let us first examine the $1D$ case , where (\ref{mixmax uni}) reduces to the familiar multiplicative congruential sequence \footnote {For an exposition of the continuous Serial Correlation in the most general case of a mixed congruential sequence $x_{n+1}=(ax_n+c)\bmod m$, see \cite{Coveyou:1960}.} \cite{Lehmer:1951}:
$$s(x)=(ax)\bmod 1.$$
The variance reduces to,
\beqa
\var{\{u\}} \rightarrow \int_0^1 x^2 \,\mathrm{d}x  - \left( \int_0^1 x \,\mathrm{d}x \right)^2 = {1\over 3} - {1\over 4} 
= {1\over 12}. 
\eeqa
The lag-s covariance between the generated points
is straightforward if we take into account the 
recursive relation of the multiplicative congruential sequence
\cite{Knuth:1981},
\beqa
s^{s}(x)=(bx)\bmod1,~~~b\equiv a^{s}
\eeqa
and the Fourier expansion of the $\mod$ function \cite{Titchmarsh:1986},
\beqa\label{mod fourier}
x\bmod 1={1\over 2}-{1\over \pi}\sum_{m=1}^{\infty}{1\over m}\sin  (2\pi mx),~~~~x\in \Re.
\eeqa
We get,
\beqa
\cov {\{u,u^s\}} &\rightarrow&
\int_{0}^{1}xs^{s}(x)\,\mathrm{d}x-\left( \int_0^1 x \,\mathrm{d}x \right)^2
=\int_{0}^{1}x[(bx)\bmod 1]\,\mathrm{d}x-{1\over 4}=\nn \\
&=&\int_{0}^{1}x\left({1\over 2}-{1\over\pi}\sum_{k=1}^{\infty}{\sin (2\pi kbx)\over k}\right)-{1\over 4}=\nn \\
&=&-{1\over \pi}\sum_{k=1}^{\infty}{1\over k}\int_{0}^{1}\mathrm{d}x~x\sin  (2\pi kbx)=
{1\over 2\pi^2 b}\sum_{k=1}^{\infty}{1\over k^2}={1\over 12b}={1\over 12a^{s}}.
\eeqa
We conclude that the lag-s correlation coefficient reduces to:
\beqa
C^{s}= { { \cov \{u,u^s\} }  \over { \var \{u\} } }
\rightarrow {1\over a^{s}}=\mathrm{e}^{-s\ln a}.
\eeqa

Finally, we can introduce the time-decay constant $\tau$ of the auto-correlations and write,
\be
C^s = C_0 ~ \mathrm{e}^{- s/\tau},
\ee
where in the $1D$ case, $\tau=1/ \ln a$.

\subsubsection {MIXMAX}
We shall, next, proceed to the calculation of the lag-s correlation coefficient between
arbitrary pairs of vector components as they are generated through the MIXMAX sequence
(\ref {mixmax uni}).The lag-s vector  can be written as follows,
\beqa \label {mixmax s-lag}
u_{i}(t+s)= \sum_{j=1}^{N}A^s_{ij}u_{j}(t)~~  \bmod 1.
\eeqa

If the period $q$ is very large we can approximate the summations, involved in the definitions
of $\var{\{u_{i}\}}$, $\cov{\{u_{i},u_{j}^s\}}$, by integrations substituting: 
\beqa\label {Ndim limit}
{1\over q}\sum_{t=0}^{q-1}\rightarrow
\int_{0}^{1}\cdots\int_{0}^{1}\Big(\prod_{j=1}^{N}\,\mathrm{d}x^{j}\Big)
\eeqa
For the variance we have,
\beqa
\var\{u_{i}\}&=&{1\over q}\sum_{t=0}^{q-1}u_{i}^{2}(t)-
\left(
{1\over q}\sum_{t}^{q-1}u_{i}(t)
\right)^2
\rightarrow\nn \\
&\rightarrow&
\int_{0}^{1}\cdots\int_{0}^{1}\Big(\prod_{j=1}^{N}\,\mathrm{d}x^{j}\Big)x_{i}^{2}-
\bigg\{\int_{0}^{1}\cdots\int_{0}^{1}\Big(\prod_{j=1}^{N}\,\mathrm{d}x^{j}\Big)x_{i}\bigg\}^2=\nn \\
&=&{1\over 3}-{1\over 4}={1\over 12}.
\eeqa
The calculation of the covariance is a little bit more involved:
\beqa
\cov\{u_{i},u_{j}^{s}\}&=&
{1\over q}\sum_{t=0}^{q-1}u_{i}(t)u_{j}(t+s)-{1\over q^2}
\sum_{t,r=0}^{q-1}u_{i}(t)u_{j}(r)=\nn \\
&=&{1\over q}\sum_{t=0}^{q-1}\bigg\{
u_{i}(t)\bigg[
{1\over 2}-{1\over \pi}\bigg(\sum_{m=1}^{\infty}{1\over m}\sin  \Big(2\pi m\sum_{k=1}^{N}A^s_{jk}u_{k}(t)\Big)
\bigg)\bigg]\bigg\}-{1\over q^2}\sum_{t,r=0}^{q-1}u_{i}(t)u_{j}(r)\rightarrow \nn \\
&\rightarrow& {1\over 4}-{1\over \pi}\sum_{m=1}^{\infty}{1\over m}
\int_{0}^{1}\cdots\int_{0}^{1}\Big(\prod_{l=1}^{N}\mathrm{d}x^{l}\Big)x_{i}\bigg[\sin  \Big(2\pi m\sum_{k=1}^{N}A^s_{jk}x_{k}\Big)\bigg]-{1\over 4}=\nn \\
&=&-{1\over \pi}\sum_{m=1}^{\infty}{1\over m}
\int_{0}^{1}\cdots\int_{0}^{1}\Big(\prod_{l=1}^{N}\mathrm{d}x^{l}\Big)x_{i}\bigg[
\sin\Big(2\pi m A^{s}_{ji}x_{i}\Big)\cos\Big(2\pi m\sum_{\substack{k=1\\ k\ne i}}^{N}A^s_{jk}x_{k}\Big)+\nn \\
&&~~~~~~~~~~~~~~~~~~~~~~~~~~~~~~~~~~~~~~~~~~~~~~~~~~
\cos\Big(2\pi m A^{s}_{ji}x_{i}\Big)\sin  \Big(2\pi m\sum_{\substack{k=1\\ k\ne i}}^{N}A^s_{jk}x_{k}\Big)\bigg]=\nn \\
&=&{1\over \pi}\sum_{m=1}^{\infty}{1\over m}
{1\over 2\pi m A_{ji}}\int_{0}^{1}\cdots\int_{0}^{1}\Big(\prod_{\substack{l=1\\ l\ne i}}^{N}\mathrm{d}x^{l}\Big)
\cos\Big(2\pi m\sum_{\substack{k=1\\ k\ne i}}^{N}A^s_{jk}x_{k}\Big)=0\nn \\
\eeqa
In the second line we used the relations (\ref{mixmax s-lag}), (\ref{mod fourier}) and in the third
(\ref{Ndim limit}). At the following steps of the calculation we used the basic 
trigonometrical identities, isolating the suitable integration variables. 

We conclude that the correlation matrix actually vanishes:
\beqa
&&\cov\{u_{i},u_{j}^{s}\} \rightarrow 0\nn \\
&&C^{s}_{ij} \rightarrow 0.
\eeqa

\subsection{Higher order correlations}
Since the lowest order correlation vanishes, which is very encouraging, we look for some other manifestation of the deterministic algorithm for generating the pseudo-random vectors. Without derivation we just give here the result of our investigation. As it turns out, the lowest order correlation which is nonzero for all elements is the following:
\be
\la u_i(t) u_j(t+1) u_k(t+1) u_l(t+1) \ra = \delta_{i,j} \delta_{j,k-1} \delta_{j,l-2}
\ee
which means that prediction of the MIXMAX sequence is difficult but post-diction is somewhat easier - there exist simple linear relations between an element of a vector at time $t$ and three consecutive elements of the subsequent vector at $t+1$.

\section{The Spectral Test for MIXMAX}
We consider the vectors of the sequence (\ref{mixmax uni}) each paired up with its successor and want to study the joint distribution. The paired vectors can be put into a $2N$ dimensional vector space, and typically lie on a set of parallel hyperplanes.  It makes sense to adopt the definition of the spectral index as given by Knuth
\cite{Knuth:1981}, as the inverse of the maximal distance between the set of hyperplanes covering all the points. 

If the entries of the matrix $A$ are all small integers, the problem of finding the smallest set of hyperplanes is not too difficult. One should keep in mind that we are working in a very large dimension and so the resulting limitation on the accuracy of any Monte-Carlo integration is negligible. Nevertheless, the spectral index in small dimensions is a very well established goodness criterion for linear RNGs. Here we would like to extend the definition of the spectral index to the matrix recursion.

\subsection {The case $N=2$}
To illustrate, we consider first the case $N=2$ where the matrix $A$ is
 \beqa
 A&=& \left(   \begin{matrix}
      2 & 1 \\
      1 & 1 \\
   \end{matrix} \right),\nn \\
\eeqa
so that the relation between the initial vector $x$ and the subsequent vector $y$ is:
\beqa\label{2D  recursion}
y_1 = 2 x_1 + x_2 \bmod 1\nn\\
y_2 = x_1 + x_2   \bmod 1\nn\\
~~~~ x_1, x_2, y_1,y_2 \in [0,1)
\eeqa
These equations define a hyperplane in the four-dimensional space of $(x_1, x_2, y_1,y_2)$. The hyperplane in this case is of course simply a 2-plane since it is defined by the equations as the intersection of two 3-planes. The complication lies in the fact that this hyperplane will enter and exit the 4-torus multiple times, while these disjoint sheets remain parallel to each other. Therefore we now turn to studying the geometric arrangement of these sheets and defining some precise measure of the maximal distance between the parallel 3-planes that include them. 

The wrapping over the 4-torus can be taken into account by noting that the wrapped sections of the 2-plane are described by  the equations:
\beqa\label {2D recursion ws}
y_1 = 2 x_1 + x_2 - w_1\nn\\
y_2 = x_1 + x_2  - w_2,
\eeqa
where $w_1, w_2$ are some integers. The possible choices for the values of $w_1, w_2$ are restricted by the requirement that the generated
vectors lie inside the $4D$ unit hypercube, i.e.
$x_1, x_2, y_1, y_2 \in [0,1)$:
\beqa
\left\{
\begin{array}{l l l}
w_1=0, 1, 2& \\
w_2= 0, 1& \\
w_1\ge w_2 & \\
y_1=y_2+x_1-(w_1-w_2)\ge 0 \Rightarrow k=w_1-w_2=0, 1
\end{array} \right.
\eeqa
Hence, we are left with four $2D$ hyperplanes, corresponding to the pairs: 
\[
(w_1,w_2)=\{(0,0), (1,0), (1,1), (2,1)\}
\]

The vectors perpendicular to each of the $3$-hyperplanes defined separately by each of 
the two equations of (\ref{2D recursion}) are:
\beqa
\hat{U}_{1}&=&{1\over \sqrt{6}}(2, 1, -1, 0)\nn \\
\hat{U}_{2}&=&{1\over \sqrt3}(1, 1, 0, -1)\nn \\
\eeqa
Implementing the typical Gram-Schmidt process we can get an orthonormal base
on this $2D$ subspace.
\beqa
\hat{E}_{1}&=&{1\over \sqrt{3}}(1, 0, -1, 1)\nn \\
\hat{E}_{2}&=&{1\over \sqrt3}(1, 1, 0, -1)\nn \\
\eeqa
A general linear combination of the above orthonormal vectors will be perpendicular to any of the four
2-planes defined by (\ref{2D recursion ws}). However, we are looking for those linear combinations which 
correspond to configurations of equidistant parallel 3-planes which cover all the 2-planes.
To find these possible configurations, we note that the generated MIXMAX vectors should belong 
to the family of the $2D$ hyperplanes defined by:
\[
\vec{X}=(x_1, x_2, 2x_1+x_2-w_1, x_1+x_2-w_2)
\]
Next, we consider the unit vector 
\[
\vec{E}=\cos\phi \hat{E}_{1}+ \sin\phi \hat{E}_{2}
\]
and demand that it be
perpendicular to all the equidistant parallel $3D$ hyperplanes which span the generated points. 
The possible configurations that correspond to those 3-planes are given by the 
allowed values for the angle $\phi$.
To find them we denote that each parallel plane can be parametrised by $R_{w_{1}w_2}$, such that:
\beqa
\hat{E} \cdot \vec{X}=R_{w_1w_2}
\eeqa

Substituting we get,
\beqa
R_{w_1w_2}=(w_1-w_2)R_{10}+w_2R_{11},
\eeqa
where
\beqa
R_{10}&=&{1\over \sqrt3}\cos\phi \nn \\
R_{11}&=&{1\over \sqrt3}\sin\phi
\eeqa
and $R_{w_1w_2}$ is the perpendicular distance between the zero point $X_{0}=(0, 0, 0, 0)$ and the
surface $(w_1,w_2)$. We observe that the distance $R_{21}$ is a linear combination of the distances
$R_{10}$ and $R_{11}$. This will help us proceed in an exhaustive analysis of the possible cases 
after which we'll be able to extract the configuration (value of $\phi$) of parallel $3D$ hyperplanes which gives the maximum distance between them.

Since we require that the distances between adjacent planes be equal, we should have,
\[
R_{w_1w_2}=n_{w_1w_2}\lambda,
\]
where $n_{w_1w_2}$ a positive integer.
In particular, we have: $n_{21}=n_{10}+n_{11}$,
which leads us to five possible choices:
\begin{itemize}
\item {$n_{11}>n_{10}~~$} 

We should have four distinct parallel hypersurfaces:
\beqa
~~~~~~~~~~~~~~~~R_{00}&=&0\nn \\
n_{10}=1,~~~~~~R_{10}&=&\lambda\nn \\
n_{11}=2,~~~~~~R_{11}&=&2\lambda\nn \\
n_{21}=3,~~~~~~R_{21}&=&3\lambda\nn
\eeqa
Hence, $\tan\phi=2$, or $\lambda={1\over\sqrt{15}}$.

\item {$n_{10}>n_{11}~~$}

We should have four distinct parallel hypersurfaces:
\beqa
~~~~~~~~~~~~~~~~R_{00}&=&0\nn \\
n_{11}=1,~~~~~~R_{11}&=&\lambda\nn \\
n_{10}=2,~~~~~~R_{10}&=&2\lambda\nn \\
n_{21}=3,~~~~~~R_{21}&=&3\lambda\nn
\eeqa
Hence, $\tan\phi=1/2$, or $\lambda={1\over\sqrt{15}}$.

\item {$n_{10}=n_{11}~~$}

We should have three distinct parallel hypersurfaces:
\beqa
~~~~~~~~~~~~~~~~R_{00}&=&0\nn \\
n_{10}=1,~~~~~~R_{10}&=&\lambda\nn \\
n_{11}=1,~~~~~~R_{11}&=&\lambda\nn \\
n_{21}=2,~~~~~~R_{21}&=&2\lambda\nn
\eeqa
Hence, $\tan\phi=1$, or $\lambda={1\over\sqrt{6}}$.

\item {$n_{10}=0~~$}

We should have two distinct parallel hypersurfaces:
\beqa
~~~~~~~~~~~~~~~~R_{00}&=&0\nn \\
n_{10}=0,~~~~~~R_{10}&=&0\nn \\
n_{11}=1,~~~~~~R_{11}&=&\lambda\nn \\
n_{21}=1,~~~~~~R_{21}&=&\lambda\nn
\eeqa
Hence, $\phi=\pi/2$, or $\lambda={1\over\sqrt{3}}$.

\item {$n_{11}=0~~$}

Again we should have two distinct parallel hypersurfaces:
\beqa
~~~~~~~~~~~~~~~~R_{00}&=&0\nn \\
n_{11}=0,~~~~~~R_{11}&=&0\nn \\
n_{10}=1,~~~~~~R_{10}&=&\lambda\nn \\
n_{21}=1,~~~~~~R_{21}&=&\lambda\nn
\eeqa
Hence, $\phi=0$, or $\lambda={1\over\sqrt{3}}$.

\end {itemize}

\begin{figure}[hb]
\centering
\includegraphics [width=110mm, height=100mm] {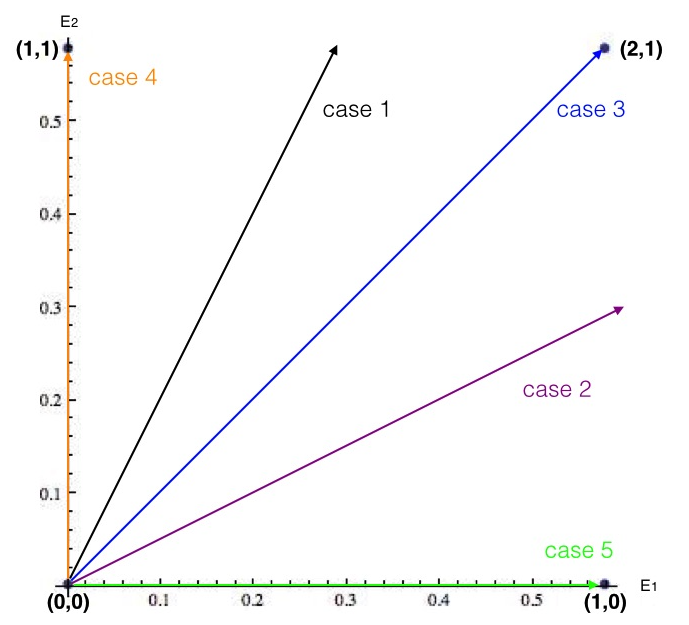}
\caption{\footnotesize{The bold points, labeled by pairs of ($w_1,w_2$) are the locations of the 2-planes in the intersection by the 2-plane spanned by $E_1,E_2$}. \label{fig:section}} 
\end{figure}

An illustrative geometrical representation of the 
five analysed cases is given in figure \ref{fig:section} where the four $2D$ hypersurfaces 
are projected as points (point projected 2-planes) and the parallel $3D$ hypersurfaces as lines 
(line projected 3-planes) perpendicular to the 5 different $\hat{E}$ vectors, intersecting one or more of the point projected 2-planes, in the $2D$ vector space of the $\hat{E}s$.

In case 1 (black line), the slope of the corresponding $\hat{E}$ vector is $\phi=\arctan{2}$.
We can view the four 3-planes as the perpendicular to the 
$\hat{E}$ vector, parallel and equidistant lines, each of which intersects a point projected 2-plane. 
The distance between them is
$1\over{\sqrt{15}}$.
Similarly in case 2 (purple line), the slope of the $\hat{E}$ is $\phi=\arctan({1\over 2})$ and the
four parallel 3-planes are projected as four, perpendicular to the 
to the $\hat{E}$ vector, lines that intersect the four point projected 2-planes.
The remaining three cases are analogously represented, the only difference being that each of the 
3-planes lines can intersect more than one 2-plane points. Indeed, in the third case the second 
3-plane line intersects two 2-plane points, while in the last two cases we are left with two
3-plane lines, each of which intersects two 2-plane points. 

In the last two cases we get the greatest distance between the parallel planes ${1\over \sqrt3}$, which leads us to
the conclusion that the spectral index is $\nu=\sqrt3$. 

\subsection {The general case}
Let us consider an arbitrary value for $N$ and try to express our results in closed and
compact forms.
The general form of the MIXMAX matrix is,
 \beqa
 A&=& \left(   \begin{matrix}
    2 & 3 & 4 & \dots & N-4 & N-3 & N-2 & N-1 & N & 1 \\
    1 & 2 & 3 & \dots & N-5 & N-4 & N-3 & N-2 & N-1 & 1 \\
    1 & 1 & 2 & \dots & N-6 & N-5 & N-4 & N-3 & N-2 & 1 \\
    \dots \\
    \dots \\
    \dots\\
    1 & 1 & 1 & \dots & 1 & 1 & 2 & 3 & 4 & 1 \\
    1 & 1 & 1 & \dots & 1 & 1 & 1 & 2 & 3 & 1 \\
    1 & 1 & 1 & \dots & 1 & 1 & 1 & 1 & 2 & 1 \\
    1 & 1 & 1 & \dots & 1 & 1 & 1 & 1 & 1 & 1 \\
   \end{matrix} \right),\\
   y_{i}&=&\left(\sum_{j=1}^{N}A_{ij}x_j\right)\bmod 1,
~~~\text{where} ~~x_1, x_2, x_3, \dots, x_N \in [0,1).\nn
\eeqa

The $N$-dimensional vectors will be paired up with their successors, thus forming a $2ND$ vector space.
The successors are given by the $N$ families of equations below, each of which corresponds to a set of $(2N-1)$-dimensional surfaces embedded in the $2N$-dimensional unit hypercube: 
\beqa\label{matrix equations N}
y_1&=&2x_1+3x_2+4x_3+\dots+(N-3)x_{N-4}+(N-2)x_{N-3}+(N-1)x_{N-2}+
Nx_{N-1}+x_{N}-w_1\nn \\
y_2&=&x_1+2x_2+3x_3+\dots+(N-4)x_{N-4}+(N-3)x_{N-3}+(N-2)x_{N-2}+
(N-1)x_{N-1}+x_{N}-w_2\nn \\
y_3&=&x_1+x_2+2x_3+\dots+(N-5)x_{N-4}+(N-4)x_{N-3}+(N-3)x_{N-2}+
(N-2)x_{N-1}+x_{N}-w_3\nn \\
&\dots&\nn \\
&\dots&\nn \\
y_{N-3}&=&x_1+x_2+x_3+\dots+x_{N-4}+2x_{N-3}+3x_{N-2}+
4x_{N-1}+x_{N}-w_{N-3}\nn \\
y_{N-2}&=&x_1+x_2+x_3+\dots+x_{N-4}+x_{N-3}+2x_{N-2}+
3x_{N-1}+x_{N}-w_{N-2}\nn \\
y_{N-1}&=&x_1+x_2+x_3+\dots+x_{N-4}+x_{N-3}+x_{N-2}+
2x_{N-1}+x_{N}-w_{N-1}\nn \\
y_{N}&=&x_1+x_2+x_3+\dots+x_{N-4}+x_{N-3}+x_{N-2}+
x_{N-1}+x_{N}-w_{N}.\nn \\
\eeqa

The vectors perpendicular to each of the $N$ families of the $(2N-1)$-dimensional hypersurfaces,
defined by (\ref{matrix equations N}), are:
\beqa
\hat{U}_{1}&=&{1\over \sqrt{1+{1\over 6}N(N+1)(2N+1)}}(2, 3, 4, \dots, N-2, N-1, N, 1; -1, 0, 0, 0, \dots, 0)\nn \\
\hat{U}_{2}&=&{1\over \sqrt{2+{1\over 6}(N-1)N(2N-1)}}(1, 2, 3, \dots, N-2, N-1, 1; 0, -1, 0, 0, \dots, 0)\nn \\
\hat{U}_{3}&=&{1\over \sqrt{3+{1\over 6}(N-2)(N-1)(2N-3)}}(1, 1, 2, \dots, N-2, 1; 0, 0, -1, 0, 0, \dots, 0)\nn \\
\dots\nn \\
\dots\nn \\
\dots\nn \\
\hat{U}_{N-3}&=&{1\over \sqrt{(N-3)+1+2^2+3^3+4^2}}
(1, 1, 1, \dots, 1, 2, 3, 4, 1; 0, 0, \dots, -1, 0, 0, 0)\nn \\
\hat{U}_{N-2}&=&{1\over \sqrt{(N-2)+1+2^2+3^2}}(1, 1, 1, \dots, 1, 2, 3, 1; 0, \dots, 0, 0, -1, 0, 0)\nn \\
\hat{U}_{N-1}&=&{1\over \sqrt{(N-1)+1+2^2}}(1, 1, 1, \dots, 1, 2, 1; 0, 0, \dots, 0, 0, -1, 0)\nn \\
\hat{U}_{N}&=&{1\over \sqrt{N+1}}(1, 1, 1, \dots, 1, 1; 0, 0, \dots, 0, 0, -1)\nn \\
\eeqa

As in the $N=2$ case, a general linear combination of above vectors will be
perpendicular to any of the N-planes defined by (\ref{matrix equations N}).
We should proceed to find those linear combinations which 
correspond to configurations of equidistant parallel $(2N-1)$-planes which cover all the $N$-planes.
However, it is not easy to implement the Gram Schmidt orthonormalization process in all its 
magnitude and get a general expression for the orthonormal basis of this $N$-dimensional  subspace. 
Fortunately, in our analysis what we actually need are just the last two
orthonormal vectors, which are easy to get. Implementing the orthonormalization process with $\hat{U}_{N}$ as the reference vector, one should get:
\beqa\label{orthonormal, N}
\dots\nn \\
\dots\nn \\
\hat{E}_{N-1}&=&{1\over \sqrt{3}}(0, 0, 0, \dots, 1, 0; 0, 0, \dots, 0,  -1, 1)\nn \\
\hat{E}_{N}&=&{1\over \sqrt{N+1}}(1, 1, 1, \dots, 1, 1; 0, 0, \dots, 0, 0, -1)
\eeqa

Since we are restricted inside the $2N$-dimensional  hypercube, i.e.
$x_1, x_2, \dots, x_N, y_1, y_2, \dots, y_N \in [0,1)$, we should have: 
\beqa
\left\{
\begin{array}{l l | l}
w_1=0, 1, \dots, {N(N+1)\over 2}-1& \\
\\
w_2= 0, 1, \dots, {(N-1)N\over 2}& \\
\\
w_3=0, 1, \dots, {(N-2)(N-1)\over 2}+1& \\
\dots\\
\dots\\
w_{N-2}=0, 1, \dots, N+2& \\
w_{N-1}=0, 1, \dots, N& \\
w_{N}=0, 1, \dots, N-1& \\
w_1\ge w_2 \ge \dots\ge w_{N-1}\ge w_{N}
\end{array} \right.
\eeqa

In addition,
\begin {itemize}
\item {$w_1=w_2+k_1$}
\beqa
y_1&=&2x_1+3x_2+4x_3+\dots+(N-3)x_{N-4}+(N-2)x_{N-3}+(N-1)x_{N-2}+
Nx_{N-1}+x_{N}-w_1\nn \\
&=&y_2+x_1+x_2+x_3+\dots x_{N-4}+x_{N-3}+x_{N-2}+x_{N-1}-k_1\ge 0 \Rightarrow \nn \\
k_1&=&0, 1, 2, \dots, N-1 
\eeqa
\item {$w_2=w_3+k_2$}
\beqa
y_2&=&x_1+2x_2+3x_3+\dots+(N-3)x_{N-3}+(N-2)x_{N-2}+
(N-1)x_{N-1}+x_{N}-w_2\nn \\
&=&y_3+x_2+x_3+\dots+x_{N-3}+x_{N-2}+x_{N-1} -k_2\ge 0 \Rightarrow \nn \\
k_2&=&0, 1, 2, \dots, N-2
\eeqa
\dots \nn \\
\dots \nn \\
\dots
\item {$w_{N-2}=w_{N-1}+k_{N-2}$}
\beqa
y_{N-2}&=&x_1+x_2+x_3+\dots+x_{N-4}+x_{N-3}+2x_{N-2}+
3x_{N-1}+x_{N}-w_{N-2}\nn \\
&=&y_{N-1}+x_{N-2}+x_{N-1}-k_{N-2}\ge 0  \Rightarrow \nn \\
k_{N-2}&=&0, 1, 2
\eeqa
\item {$w_{N-1}=w_{N}+k_{N-1}$}
\beqa
y_{N-1}&=&x_1+x_2+x_3+\dots+x_{N-3}+x_{N-2}+
2x_{N-1}+x_{N}-w_{N-1}\nn \\
&=&y_{N}+x_{N-1}-k_{N-1}\ge 0  \Rightarrow \nn \\
k_{N-1}&=&0, 1
\eeqa
\end {itemize}
With the above restrictions on the values of the possible combinations of the ordered set of
$(w_1, w_2, w_3, \dots, w_N)$,
it is not hard to see that we are left with  $N!\cdot N$ surfaces.

The generated MIXMAX vectors should belong to the $N!\cdot N$, $N$-dimensional  hyperplanes which result by intersecting the families of the $(2N-1)$-dimensional  hypersurfaces (\ref{matrix equations N}),
\beqa\label{X, N}
\vec{X}&=&\Big(x_1,~ x_2, \dots, x_{N-1},~ x_N; \nn \\ 
&&~~~2x_1+3x_2+4x_3+\dots+(N-2)x_{N-3}+(N-1)x_{N-2}+
Nx_{N-1}+x_{N}-w_1,\nn \\
&&~~~x_1+2x_2+3x_3+\dots+(N-3)x_{N-3}+(N-2)x_{N-2}+
(N-1)x_{N-1}+x_{N}-w_2,\nn \\ 
&&~~~\cdots ,\nn \\ 
&&~~~x_1+x_2+x_3+\dots+x_{N-4}+x_{N-3}+x_{N-2}+
2x_{N-1}+x_{N}-w_{N-1}, \nn \\
 &&~~~x_1+x_2+x_3+\dots+x_{N-4}+x_{N-3}+x_{N-2}+
x_{N-1}+x_{N}-w_{N}\Big).
\eeqa

A unit vector inside the vector space with basis $\hat{E}_{1}, \hat{E}_{2}, \dots \hat{E}_{N}$ 
can be parametrized through the $N-1$ spherical coordinate parameters,
$ \{(\theta_1, \theta_2,\dots, \theta_{N-2}, \phi): \theta_1, \theta_2, \dots,  \theta_{N-2} \in [0,\pi], ~\phi\in [0,2\pi) \}$.

\beqa\label{E, N}
\hat{E}&=&\sin\theta_1 \sin\theta_2\dots \sin\theta_{N-2}\sin\phi\hat{E}_{1}+\sin\theta_1\sin\theta_2\dots\sin\theta_{N-2}\cos\phi \hat{E}_{2}+\dots +\nn \\
&&+\sin\theta_1\sin\theta_2\cos\theta_3\hat{E}_{N-2}
+\sin\theta_1\cos\theta_2\hat{E}_{N-1}+\cos\theta_1 \hat{E}_N
\eeqa

We should demand this vector be
perpendicular to all the parallel and equidistant $(2N-1)$-dimensional  hyperplanes which span the generated points. 
Each of the parallel hyperplanes can be parametrised by its distance from the origin 
$R_{w_1w_2\dots w_{N} }$
such that,
\beqa\label{dot EX, N}
\hat{E} \cdot \vec{X}=R_{w_1w_2\dots w_N}.
\eeqa

Substituting (\ref{X, N}) and (\ref{E, N}) in (\ref{dot EX, N}) we get,
\beqa
R_{w_1w_2\dots w_N}&=&
(w_1-w_2)R_{1000...000}+(w_2-w_3)R_{1100...000}+
 (w_3-w_4)R_{1110...000}+
 \dots+\nn \\
 &&+(w_{N-1}-w_{N})R_{1111\dots 110}+w_{N} R_{1111\dots 111},
\eeqa

where,
\beqa
R_{w_1w_2\dots w_N}=R_{w_1w_2\dots w_N} (\theta_1, \theta_2, \dots, \theta_{N-2}, \phi)
\eeqa

As in the case $N=2$  the distances from the origin of all the $N!N$, $(2N-1)D$ hypersurfaces,
can be expressed as linear combinations of the above $N$ basis hypersurfaces, with
positive definite coefficients. The values of these distances depend on the choice of the $N-1$ parameters
$(\theta_1, \theta_2, \dots, \theta_{N-2}, \phi)$. 

From (\ref{orthonormal, N}) and (\ref{X, N}) we get,
\beqa
\hat{E}_{N}\cdot \vec{X}&=&{w_{N}\over \sqrt{N+1}}\nn \\
\hat{E}_{N-1}\cdot \vec{X}&=&{1\over \sqrt3}(w_{N-1}-w_{N})
\eeqa

Hence (\ref{dot EX, N}) becomes,
\beqa
R_{w_1w_2\dots w_{N} }&=&\sin\theta_1\sin\theta_2~ h(\theta_3, \dots, \theta_{N-2}, \phi, w_1, w_2, \dots, w_N)+\left({w_{N-1}-w_{N}\over \sqrt3}\right)\sin\theta_1\cos\theta_2+\nn \\
&&+
{w_{N}\over \sqrt{N+1}}\cos\theta_1.
\eeqa
What we have managed is to factorise, in each term, the dependence of 
$R_{w_1w_2\dots w_{N} }$ on the angle parameters $\theta_1, \theta_2$.

The choice of the angle parameters
$\left(\theta_1, \theta_2, \dots, \theta_{N-2}, \phi\right)=\left({\pi\over 2}, 0, \theta_3, \dots, \theta_{N-2}, \phi\right)$ 
 will force all but the $R_{1111\dots 110}$ basis hypersurfaces nullify. Indeed we get,

\beqa
R_{w_1w_2\dots w_N}\left({\pi\over 2}, 0, \theta_3, \dots, \theta_{N-2}, \phi\right)={1\over \sqrt3}(w_{N-1}-w_{N})=
{1\over \sqrt3}k_{N-1}
\eeqa
so that:
\beqa
R_{1000\dots 000}&=&R_{1100\dots 000}=\dots R_{111\dots 100}=R_{111\dots 111}=0\nn \\
R_{1111\dots 110}&=&{1\over \sqrt3}
\eeqa
For such a choice of the angles, half of the total parallel $(2N-1) D$ hypersurfaces
(the surfaces $R_{w_1w_2\dots w_{N-2}ll}$)
will coincide with the surface passing from the origin: $R_{0000\dots 000}$, and half
of them (the surfaces $R_{w_1w_2\dots w_{N-2}, l+1, l}$)
will collapse to the surface $R_{1111\dots 1110}$.

As we have seen in the case of $N=2$,  such a configuration gives us the least 
possible number of parallel planes and most probably the largest possible distance
between the adjacent ones. Thus, we have obtained an upper bound on the spectral index of $\nu=\sqrt3$ for any matrix dimension $N$.
Because this number is of order one, we can conclude from this that the equidistribution of the MIXMAX vectors in 2N dimensions is not very good. However, this by itself is not a reason for concern, because it limits the ultimate precision of almost any Monte-Carlo simulation in $2N$ dimensions to O( $\sqrt3^{-2N}$ ) which is an infinitesimally small number.

\section{Acknowledgements}
As interesting new results of L'Ecuyer, Wambergue, Bourceret 
have become available to us \cite{L?Ecuyer:2017}, we think that it is worth to share our initial investigation 
on the spectral index which has been performed by a different method.

 We thank G. Savvidy, E. Cheung, J. Hladky, J. Apostolakis, J. Harvey, L. Moneta and G. Georgiou for 
 the useful discussions. 
 
This project has received funding from the European Union's Horizon 2020 research and innovation programme under the Marie Sklodowska-Curie grant agreement no. 644121.

\end{document}